\documentstyle[twocolumn,prl,aps,psfig]{revtex}
\def\prb{Phys. Rev. B}
\def\prl{Phys. Rev. Lett.}

\def\LNiO{La$_2$NiO$_{4}$}
\def\LNiOd{La$_2$NiO$_{4+\delta}$}
\def\LSCO{La$_{2-x}$Sr$_x$CuO$_4$}
\def\LCO{La$_2$CuO$_4$}
\def\LSNiOd{La$_{2-x}$Sr$_x$NiO$_{4+\delta}$}
\def\LSNiO{La$_{1\frac{2}{3}}$Sr$_{\frac{1}{3}}$NiO$_{4}$}

\def\KNiF{K$_{2}$NiF$_{4}$}
\def\cm-1{cm$^{-1}$}

\begin{document}
\twocolumn[
\hsize\textwidth\columnwidth\hsize\csname@twocolumnfalse\endcsname
\draft

\preprint{}
\title
{Charge and Spin Dynamics of an Ordered Stripe Phase in \LSNiO \ by Raman 
Spectroscopy 
}
\author{
G.~Blumberg,$^{1,2,\dag}$ M.~V.~Klein,$^{1}$ and S-W. Cheong$^{3}$  
}
\address{
$^{1}$NSF Science and Technology Center for Superconductivity and \\
Department of Physics, 
University of Illinois at Urbana-Champaign, Urbana, IL 61801-3080\\
$^{2}$Institute of Chemical Physics and Biophysics, R\"avala 10, 
Tallinn EE0001, Estonia\\ 
$^{3}$Bell Laboratories, Lucent Technologies, Murray Hill, NJ 07974 and\\ 
Department of Physics and Astronomy, Rutgers University,
Piscataway, NJ 08855
}
\date{July 7, 1997; to appear in Phys. Rev. Lett.}
\maketitle

\begin{abstract}
For \LSNiO \ -- a commensurately doped Mott-Hubbard system --   
charge- and spin-ordering in a stripe phase has been 
investigated by phononic and magnetic Raman scattering. 
Formation of a superlattice and an opening of a pseudo-gap in the 
electron-hole excitation spectra as well as two types of double-spin  
excitations --  within the antiferromagnetic domain and across 
the domain wall -- are observed below the charge-ordering transition.  
The temperature dependence suggests that the spin ordering is 
driven by charge ordering and that fluctuating stripes persist above 
the ordering transition. 
\end{abstract}

\pacs{PACS numbers: 71.45.Lr, 75.30.F, 74.70.Kb, 71.38.+i, 74.72.Dn, 
78.30.Er, 74.25.Ha 
}
]
%
%\narrowtext

The problem of doped Mott-Hubbard insulators has attracted much 
attention because of its relevance to the cuprate high temperature 
superconductors. 
Recent experiments on the doped lanthanum nickelate \cite{Chen93,Tranquada94,NiO} 
and lanthanum cuprate 
\cite{TranquadaNature,NdCuO,CuO} families of materials have established 
a new type of 
real space charge and spin ordering in topological phases.
In the low-temperature stripe phase, the doped holes are concentrated in 
periodic antiphase domain walls for the intervening antiferromagnetic 
regions -- see Fig.~1. 
These results appear to be consistent with the ideas of frustrated phase 
separations driven by the electron-electron interaction \cite{Emery}. 

The stripe phase was first observed by electron diffraction 
\cite{Chen93} and by neutron scattering measurements 
\cite{Tranquada94} in the \LSNiOd \ system.  
Undoped \LNiO \ is an antiferromagnetic charge-transfer insulator 
with a O-$2p$~--~Ni-$3d$ gap of about 3.5~eV \cite{Ido91,Anisimov92}, 
twice as large as that in the isostructural \LCO. 
The ground state has spin ${\bf S}=1$ with holes localized on 
Ni$^{2+}$ ($x^{2}-y^{2}, 3z^{2}-1$) sites. 
Introduction of holes into the NiO$_{2}$ plane induces broad optical 
absorption in the infrared region below the charge-transfer gap, 
similar to the case of \LSCO, but unlike the cuprates the lanthanum nickelate 
remains insulating up to high doping levels and a 
Drude-type absorption band does not appear \cite{Ido91,Bi93}.  
It has been suggested that 
the difference from the cuprates is the larger effect of the 
electron-phonon coupling \cite{Anisimov92,Zaanen94} and stronger 
magnetic localization (less important spin fluctuations for spin-one 
system) \cite{KivelsonEmery94}. 
The doped hole is nearly entirely localized on Ni and four nearest 
neighboring in-plane oxygen atoms. 
The high-frequency vibrations of the light oxygens participate in the 
electron-phonon interaction and form a small polaron. 
The spin of the additional hole is antiparallel to that of Ni$^{2+}$ and 
leads to low spin on the doped nickel sites \cite{Anisimov92}.  

The neutron scattering measurement on \LSNiO, in 
which the number of doped holes per Ni site is 1/3, has revealed that 
the system undergoes three successive transitions associated  with 
quasi-two-dimensional commensurate charge and spin stripe ordering in 
the NiO$_{2}$ planes \cite {Lee97}. 
The low-temperature phases are suggested to be static stripe lattice 
states with quasi-long-range in-plane charge correlation (See Fig.~1). 
Previously for the system with the same 1/3 doping 
anomalies associated with the charge ordering 
transition at $T_{co} \approx 240$~K have been observed 
by electronic diffraction measurement \cite{Chen93} as well as 
in resistivity, magnetic susceptibility, sound velocity and specific heat 
\cite{Cheong94,Ramirez96}.  
In a recent optical study, Katsufuji {\em et al.}  \cite{Katsufuji96} 
observed that 
when temperature is decreased down to $T_{co}$ the low-energy 
spectral weight below $\sim$0.4~eV becomes suppressed and is transferred 
to higher energy up to $\sim$2~eV.   
For $T < T_{co}$ spectral weight below $\sim$0.26~eV becomes strongly 
reduced (A charge gap $2\Delta(T)$ opens up.), and the missing spectral 
weight is also distributed over the higher energy region up to 
$\sim$2~eV.   
In this paper we study both charge and spin ordering by Raman 
spectroscopy. 
The data were taken on a \LSNiO \ single crystal grown as described 
in Ref. \cite{Lee97}. 

The Raman spectra were measured in the back-scattering geometry from the 
sample mounted in a continuous helium flow optical cryostat (Oxford Instruments).  
Throughout this study, we used linearly polarized 6471~\AA \ 
excitation from a Kr$^+$ laser. 
To reduce the heating by laser illumination we used about 5~mW of 
incident laser power focused onto a 50~$\mu$m diameter spot on the 
$ab$-plane of the as grown mirror-like crystal surface. 
The incident photons were polarized along the direction $45^{\circ}$ from 
the Ni-O bond.  
The crystallographic orientation was determined by X-ray Laue diffraction. 
The scattering photons polarized either parallel 
($x^{\prime} x^{\prime}$ geometry) or perpendicular 
($x^{\prime}y^{\prime}$ geometry) 
to the incident photons were collected and analyzed by a triple 
grating spectrometer with a liquid nitrogen cooled CCD detector. 
With assumed approximate $D_{4h}$ symmetry, these two geometries 
correspond to spectra of $A_{1g} + B_{2g}$ and $B_{1g} + A_{2g}$ 
symmetry, respectively. 
The $B_{2g}$ and $A_{2g}$ components of the spectra ($xy$ geometry) are 
found to be quite weak. 
Spectra were corrected for the spectral response of the spectrometer and 
the detector. 

Figure 2 shows the Raman spectra for two polarizations as a function of 
temperature. 

\emph{Continuum.} 
At high temperatures spectra in both geometries show strong continuum 
intensity. 
In metals electronic Raman scattering by charge fluctuations 
arises from electron-hole excitations near the Fermi energy.     
For a normal metal within the Fermi liquid model the electronic 
Raman scattering has finite intensity only at very low frequencies. 
For strongly correlated systems, incoherent quasi-particle scattering 
leads to finite Raman intensity over a broad region of frequency 
\cite{Varma89,Shastry}, 
and the Raman intensity can be used as a measure of the incoherent 
quasi-particle scattering. 
At high temperatures the strong continuum scattering starts from very low 
frequencies, indicating gapless electron-hole excitations of the doped \LSNiO. 
The low-frequency portion of the continuum intensity reduces with 
temperature reduction to 150~K indicating, 
in agreement with resistivity and optical conductivity 
studies \cite{Katsufuji96}, that a pseudo-gap opens up in the low-frequency 
electronic excitation spectra. 
 
\emph{Phonons.} 
Temperature dependence of optical-phonon structures provides evidence 
for the charge ordering transition at $\sim$~240~K. 

At room temperature \LSNiO \ has the tetragonal \KNiF \ structure with 
the space group $I4/mmm$ \cite{Han95}, 
in which there should be four Raman-active phonon modes; 
Ni-O(2) and O(2)-La stretching along $c$-axis modes with the  
$A_{1g}$ symmetry and the $E_{g}$ symmetry 
O(2) and La vibrations along the $a$- or $b$-axis 
\cite{Bates89,Burns90,Anders91,Udagawa93}.  
For the $A_{1g}$ symmetry ($x^{\prime} x^{\prime}$ geometry) the most 
intense peak at around 500~\cm-1 has been assigned to the oxygen 
stretching mode. 
This mode shows hardening with Sr doping \cite{Udagawa93}. 
We attribute the peaks around 140~\cm-1 to the La stretching mode. 
Above the charge ordering temperature all the observed modes are weak;  
the 140 and especially 500~\cm-1 modes are broad, indicating 
strong polaronic effects \cite{Anisimov92} and inhomogeneous charge 
distribution.   

Conspicuous changes in phononic spectra are observed below $T_{co}$. 
The charge ordering gives rise to formation of a superlattice, 
triples the unit cell size and lowers the crystal symmetry (See Fig.~1). 
It leads to folding of phonon-dispersion branches and the 
appearance of new $\Gamma$-point Raman-active modes. 
In agreement with this picture, spectra exhibit a number of new sharp phononic 
modes at $T < T_{co}$ \cite{assignment}.  
The $A_{1g}$ modes observed above $T_{co}$ show significant 
sharpening and intensity enhancement at low temperatures. 
The Ni-O(2) stretching mode splits below $T_{co}$ into two 
narrow peaks at 474 and 515~\cm-1 (Fig.~3) indicating that there are two 
nonequivalent Ni-O(2) bonds below the charge ordering temperature. 
The first might be assigned to the undoped Ni atoms between the domain 
walls, and the second to the Ni atoms on the walls where doped holes 
are tightly localized (See Fig.~1). 
Katsufuji {\em et al.} observed splitting below $T_{co}$ of the IR-active 
$E_{u}$ symmetry 355~\cm-1 in-plane Ni-O(1) bending mode \cite{Katsufuji96} 
that also can be understood in terms of two types of Ni-O(1) bonds in 
the charge-ordered phase.     

In the Fig.~4 we show the temperature dependence of the 
Ni-O(2) stretching mode Raman intensity integrated between 400 and 
640~\cm-1 above the continuum. 
Analogous temperature dependence has been observed for other phononic 
modes. 
The intensity shows strong enhancement below $T_{co}$, and its 
temperature dependence is similar to the temperature dependence of the 
gap magnitude $2\Delta(T)$ from optical measurements \cite{Katsufuji96} that has 
been suggested to be a relevant order parameter for the charge-ordered phase. 
The presented Raman spectra are excited with 1.9~eV photons, energy 
in the region where the spectral weight is transferred when the charge gap 
opens up below $T_{co}$ \cite{Katsufuji96}. 
Hence, the intermediate state of the scattering process is in resonance 
with states that are sensitive to the charge ordering transition. 
The oxygen and La phonon modes, on the other hand, modulate the 
atomic displacement to produce the commensurate charge-ordered state.  
Therefore, the resonant Raman intensity of the phononic modes that 
participate in the charge ordering transition is coupled {\it via} the 
intermediate state to the order parameter $2\Delta$. 

\emph{Spin excitations.} 
Short wavelength antiferromagnetic fluctuations, which may exist 
without long-range antiferromagnetic order, can be probed by Raman 
scattering. 
The Raman process takes place from antiferromagnetic (${\bf S}=1, 
S_{z}=\pm 1$) ground state {\it via} a photon-stimulated virtual 
charge-transfer excitation that exchanges two spins on nearest 
neighbor Ni sites in an intermediate two singlets (${\bf S}=0, S_{z}=0$) state 
(each of two exchanged spins compensates with the remaining spins on 
the same sites) \cite{Shastry}, 
and then to a final triplet (${\bf S}=1, S_{z}=0$) state. 
This process may also be described as creation of two interacting magnons. 

The undoped \LNiO \ antiferromagnetic insulator has been studied by Sugai 
{\it et al.} \cite{Sugai90}. 
The $B_{1g}$ spectra exhibits a band peaked at about 1640~cm$^{-1}$. 
This band has been assigned to scattering by two-magnons. 
For ${\bf S}=1$ system two-magnon excitation requires about $\sim 7J$ energy 
due to increase of the Heisenberg interaction  
$J \sum_{<i,j>} ({\bf S}_i \cdot {\bf S}_j - \slantfrac{1}{4})$ 
[$J$ is spin superexchange constant, 
${\bf S}_i$ is the spin on site $i$ and the summation is over near-neighbor 
Ni pairs.]. 
Thus for the antiferromagnetic insulators the two-magnon peak position 
yields an estimate of $J \approx 240$~\cm-1. 
 
For doped \LSNiO \ we have not observed a band near 1640~cm$^{-1}$ at 
any temperature. 
Instead, at low temperature the spectra in $x^{\prime}y^{\prime}$ 
polarization exhibit two bands peaked at $\sim 720$ and 
$\sim 1110$~\cm-1. 
We interpret these two bands as double-spin excitations with reduced energy due 
to reduction in the number of magnetic nearest neighbors (broken magnetic 
bonds) in the vicinity of charge domain walls \cite{superexchange}. 
The first band is double-spin excitation within the 
antiferromagnetic domain with Heisenberg magnetic energy price of 
$\sim 3J$, and the second one across the domain wall with 
$\sim 4J$ energy, respectively (See Fig. 1). 
The observed reduction of the two-magnon energy relative to the energy in 
the undoped antiferromagnetic insulators is consistent with the 
neutron scattering study of \LNiOd \ by Tranquada {\it et al.} 
\cite{Tranquada96b} where 
it was found that the spin-wave velocity for excitations propagating parallel 
to the stripes in $\delta = 0.133$ sample is $\approx 60\%$ of that in 
undoped \LNiO. 
 
The temperature dependence of the magnetic scattering intensity 
integrated above the continuum up to 1400~\cm-1 and the intensity  
portion under the second peak (integrated between 875 and 
1400~\cm-1) are shown in the Fig.~4. 
Above $T_{co}$ the magnetic excitations are overdamped; however, a weak 
peak about 600~\cm-1 is seen on the strong continuum background 
suggesting that the fluctuating stripes exist above the charge-order 
transition.  
When the charge gap $2\Delta$ opens up below $T_{co}$, the 720~\cm-1 
two-magnon band, freed from the damping, emerges from the overdamped 
excitation.  
The temperature dependence of the magnetic scattering intensity and 
shape indicates 
that the antiferromagnetic ordering within the undoped region is 
retarded relative to the charge ordering transition.  
The short range antiferromagnetic order begins to be established at perhaps 
about 200~K, a temperature where the long range static charge 
correlations have been observed by elastic neutron scattering 
\cite{Lee97}, and the quasi-one-dimensional antiferromagnetic ordering 
continues down to the lowest temperature. 
The 1110~\cm-1 band intensity indicates that the relative $\pi$-phase 
shift between the antiferromagnetic regions becomes established at 
lower temperatures. 

In summary, 
lattice and spin dynamics of \LSNiO, which undergoes charge and spin 
ordering transitions starting from 240~K, are investigated by 
Raman spectroscopy. 
In the high-temperature phase the Raman continuum demonstrates strong 
incoherent quasi-particle scattering and gapless electron-hole 
excitation. 
Below 240~K a pseudo-gap opens up in the low-frequency electronic 
excitations. 
The charge ordering gives rise to formation of a superlattice that is 
observed in splitting of optical phonon and appearance of new 
Raman-active modes. 
The resonance Raman scattering intensity of these phononic modes 
reflects  
redistribution of the electronic spectra caused by charge ordering. 
The magnetic Raman scattering shows two bands corresponding to photon 
induced double spin excitation within the antiferromagnetic domain and across 
the domain wall. 
Their temperature dependence suggests that the magnetic order is driven 
by charge order and reflects the temperature dependence of the short 
range magnetic correlations within antiferromagnetic domain as well as 
across the domain wall in the stripe phase. 
Overdamped short-range antiferromagnetic correlations are observed above 
$T_{co}$, suggesting fluctuating stripes in the high 
temperature phase. 

\bigskip
 
We acknowledge discussions with V.J.~Emery, S.A.~Kivelson, M. Salamon and 
J. Tranquada.
This work was supported by NSF grant DMR 93-20892 and cooperative agreement 
DMR 91-20000 through the STCS.

\begin{figure}
\centerline{
\psfig{figure=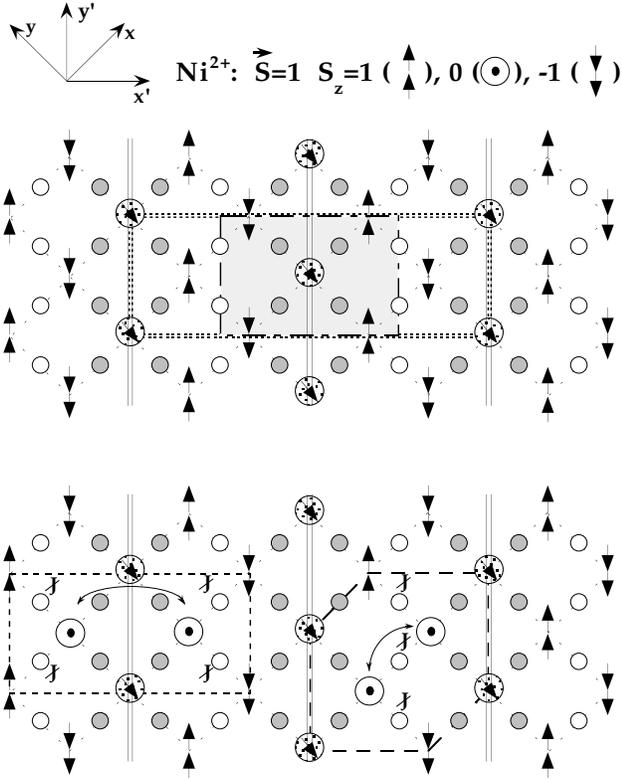,width=8.64cm,clip=}} 
\caption{ 
The low-temperature stripe phase for 1/3 doping. 
Arrows indicate uncompensated spins on Ni sites; 
circles indicate oxygen sites; 
shaded circles indicate locations of doped holes on oxygen and Ni 
sites (the orientation of remaining spin on the Ni site is frustrated). 
Dashed lines trace the bonding paths of the square lattice 
(the unit cell of the tetragonal high-temperature phase). 
Double lines indicate position of the domain walls. 
The low-temperature charge unit cell is shown by shaded area while 
double broken lines outline a magnetic unit cell. 
The lower panel illustrates two types of double-spin excitations:  
the double-spin excitation within an antiferromagnetic domain breaks 
three magnetic bonds (broken bonds are shown as 
$\not \negthinspace J$), 
while the excitation across the domain wall breaks four bonds 
\protect\cite{superexchange}. 
}
\label{Fig.1}
\end{figure}

\begin{figure}
\centerline{
\psfig{figure=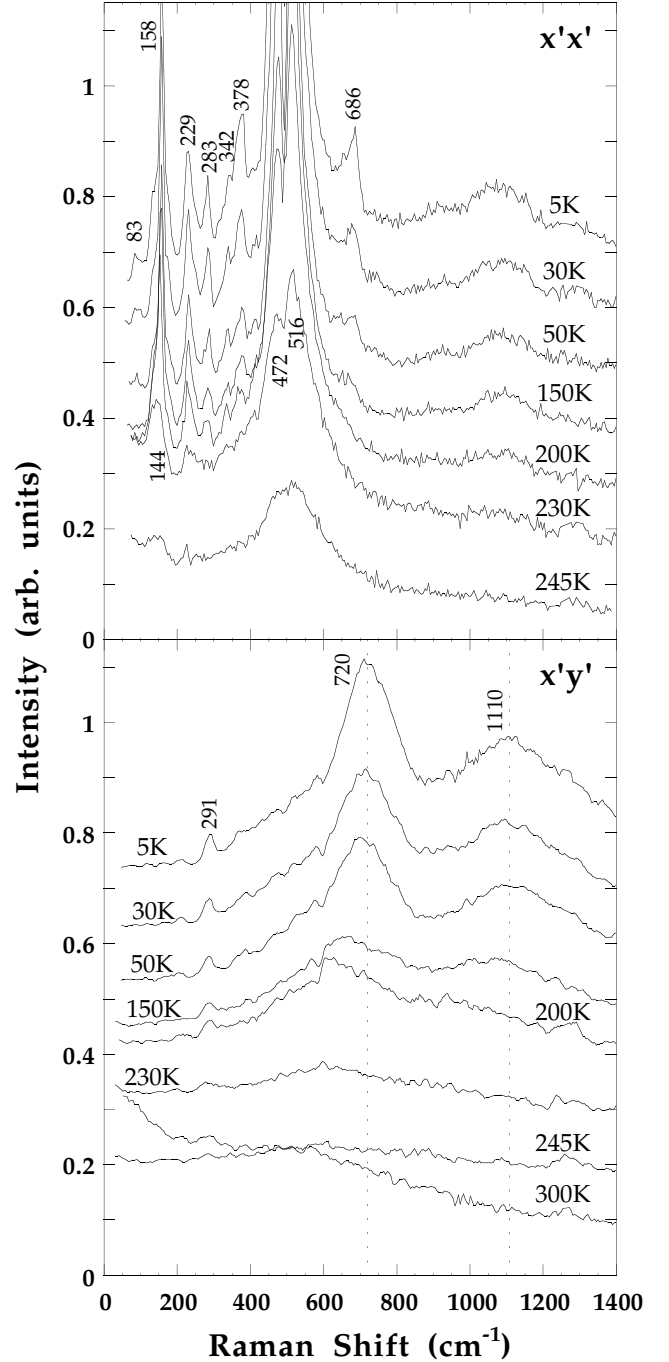,width=8.64cm,clip=}}
\caption{ 
Temperature dependent Raman scattering spectra for $x^{\prime} x^{\prime}$ 
and $x^{\prime}y^{\prime}$ polarization. 
Baselines are shifted by one tick. 
}
\label{Fig.2}
\end{figure}

\begin{figure}
\centerline{
\psfig{figure=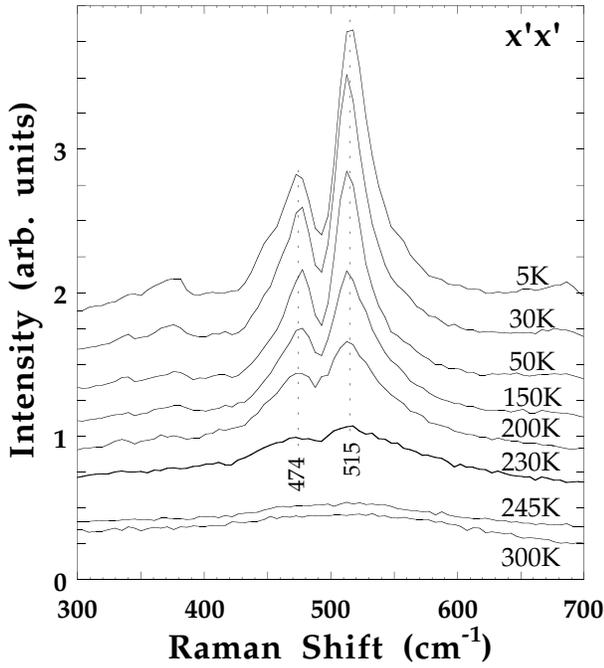,width=8.1cm,clip=}}
\caption{ 
Temperature dependent  
Raman scattering spectra for $x^{\prime} x^{\prime}$ polarization. 
Baselines are shifted by one tick. 
The 474 and 515~\cm-1 phonons are assigned to Ni-O(2) stretching modes 
on undoped and doped sites. 
}
\label{Fig.3}
\end{figure}

\begin{figure}
\centerline{
\psfig{figure=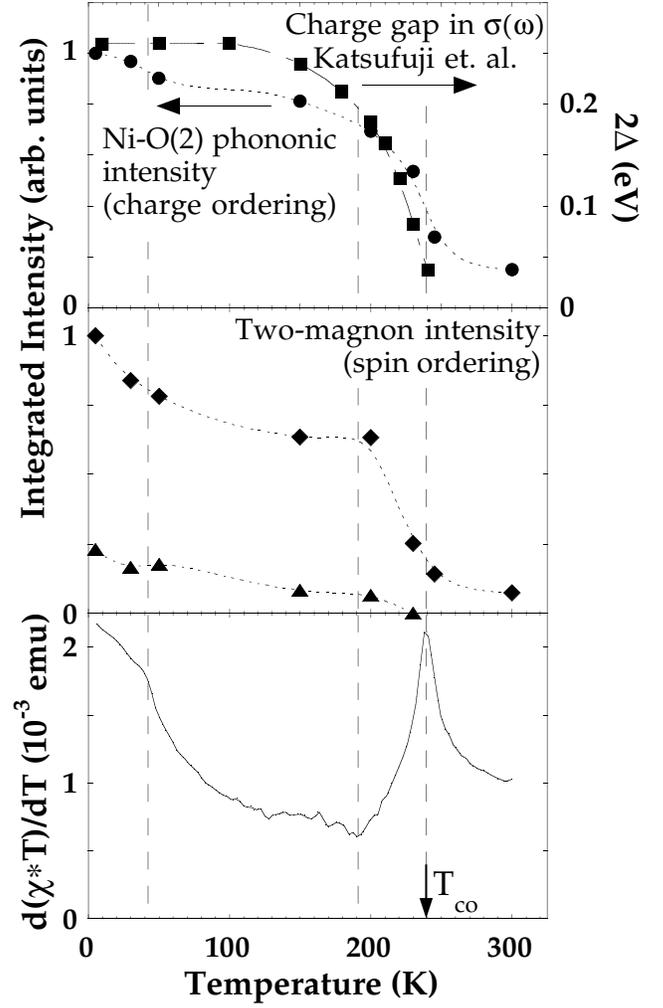,width=8.5cm,clip=}}
\caption{ 
Temperature dependence of 
the  Ni-O(2) stretching modes intensity (See Fig.~3) integrated between 
400 and 640~\cm-1 above the continuum (circles) 
shown together with the charge gap in optical conductivity  
from Ref. \protect\cite{Katsufuji96} (squares); 
the magnetic scattering intensity in both 720 and 1110~\cm-1 bands 
(See Fig.~2) integrated above the continuum up to 1400~\cm-1 (diamonds) 
and the intensity portion under the 1110~\cm-1 peak integrated between 
875 and 1400~\cm-1 (triangles); 
also anomalies in $d\,(\chi T)/d\,T$ \protect\cite{Lee97} are 
shown for comparison. 
} 
\label{Fig.4}
\end{figure}


\begin{references}
\bibitem{assignment}
The detailed assignment of these new modes is yet to be made.
\bibitem{superexchange} 
The spin superexchange across the domain wall is expected to be small.
\end{references}
\end{document}